# Turbulent dynamo in the terrestrial magnetosheath

Zoltán Vörös ®[1,2] ✉, Owen Wyn Roberts ®[3], Yasuhito Narita ®[4,5], Emiliya Yordanova ®[6], Rumi Nakamura ®[1,11], Adriana Settino ®[1,11], Daniel Schmid ®[1,11], Martin Volwerk[1,11], Cyril L. Simon Wedlund ®[1,7,11], Ali Varsani ®[1,11], Luca Sorriso-Valvo ®[8,9,11], Philippe André Bourdin ®[7,11] & Árpád Kis[2,10,11]

Dynamo action refers to energy exchange processes through which magnetic fields are generated at the expense of kinetic energy of the plasma flows. Dynamos can generate magnetic fields across scales larger or smaller than the flows themselves. Multi-scale dynamo processes underpin magnetic phenomena from planetary cores to stellar and galactic environments, while also shaping turbulent magnetic fields at smaller scales. Yet, experimental validation of dynamo action has remained largely confined to laboratories. Here we report evidence for a turbulent dynamo in the terrestrial magnetosheath. Observations reveal the predicted spatial topology of stretched and folded magnetic fields, compressive effects, and pressure anisotropy instabilities essential for magnetic field amplification. Our findings also highlight the central role of turbulent dynamos in energy conversion and structure formation within collisionless plasma turbulence. The observed energy exchange signatures indicate that the magnetosheath may serve as a natural testbed for validating dynamo theories and simulations.

Magnetic fields are present across various scales in the Universe[1], yet the mechanisms responsible for their generation or amplification remain largely untested in experimental settings. Laboratory experiments for studying the amplification of magnetic fields in vessels containing liquid metals[2], as well as laser plasma experiments[3,4], are highly useful for understanding dynamo action under various parameter regimes and boundary conditions. However, in space and astrophysical plasmas, the extent of spatial and temporal scales, as well as the complexity in terms of degrees of freedom and nonlinearity, can limit the predictive power of laboratory dynamo experiments[5,6].

Although dynamo theories and numerical simulations may offer alternatives for how magnetic fields are generated or amplified[6–8], the actual energy conversions in space and astrophysical plasmas occur in locations that are not directly accessible for measurements. For example, the geomagnetic field cannot be measured within the planetary interior where the dynamo operates; it can only be detected at the Earth's surface or beyond. In general, large-scale dynamos (LSDs) generating or amplifying magnetic fields on scales comparable to or larger than a characteristic scale of flow shears are fundamental in explaining planetary, stellar, or (extra-)galactic magnetic fields[7]. In-situ measurements of the LSD action in the solar wind indicate that local magnetic field amplification can also take place within the ejected material from the solar corona[9–11]. However, with single-point measurements in the solar wind, the three-dimensionality of fluctuations required for dynamo action becomes impossible to measure, and the temporal and spatial fluctuations cannot be fully differentiated. To overcome the limitations of single-point measurements, multi-point spatial observations are needed, which are available in the terrestrial magnetosheath. The terrestrial magnetosheath is a boundary region formed by the interaction of the solar wind with the geomagnetic field, bordered by the bow shock (outer boundary) and the magnetopause (inner boundary). It contains a turbulent, nearly collisionless, high $\beta$ (the ratio of plasma to magnetic pressure) plasma and the compressed magnetic field[12]. Typically, $\beta \gtrsim 1$ is in the magnetosheath. In this environment, four-point measurements are available from the Magnetospheric Multiscale (MMS) space mission[13]. Although MMS occasionally samples the solar wind, its plasma instruments are optimized for hot magnetospheric conditions.

[1]Space Research Institute, Austrian Academy of Sciences, Graz, Austria. [2]Institute of Earth Physics and Space Science, HUN-REN, Sopron, Hungary. [3]Department of Physics, Aberystwyth University, Aberystwyth, UK. [4]Institut für Theoretische Physik, Technische Universität Braunschweig, Braunschweig, Germany. [5]Max Planck Institute for Solar System Research, Göttingen, Germany. [6]Swedish Institute of Space Physics, Uppsala, Sweden. [7]Institute of Physics, University of Graz, Graz, Austria. [8]Space and Plasma Physics, School of Electrical Engineering and Computer Science, KTH Royal Institute of Technology, Stockholm, Sweden. [9]CNR/ISTP-Istituto per la Scienza e la Tecnologia dei Plasmi, Bari, Italy. [10]University of Sopron, Sopron, Hungary. [11]These authors contributed equally: Rumi Nakamura, Adriana Settino, Daniel Schmid, Martin Volwerk, Cyril L. Simon Wedlund, Ali Varsani, Luca Sorriso-Valvo, Philippe André Bourdin, Árpád Kis. ✉e-mail: zoltan.voeroes@oeaw.ac.at

In this work, we demonstrate that high-resolution MMS observations in the terrestrial magnetosheath capture the predicted signatures of a turbulent small-scale dynamo (SSD). We show, both statistically over an extended time interval and through targeted case studies, that stretched and folded magnetic field geometries with the expected topology, together with pressure-anisotropy instabilities supporting dynamo action[6], arise naturally in magnetosheath plasma turbulence. We further find that magnetosheath field and plasma parameters, modulated by the variable solar wind, might exhibit substantial variability.

## Results
### Observations of small-scale dynamo (SSD) in the turbulent magnetosheath

The local temporal evolution of the magnetic field at the position of the spacecraft can be described by the magnetic induction equation, which contains the magnetic field generating terms,

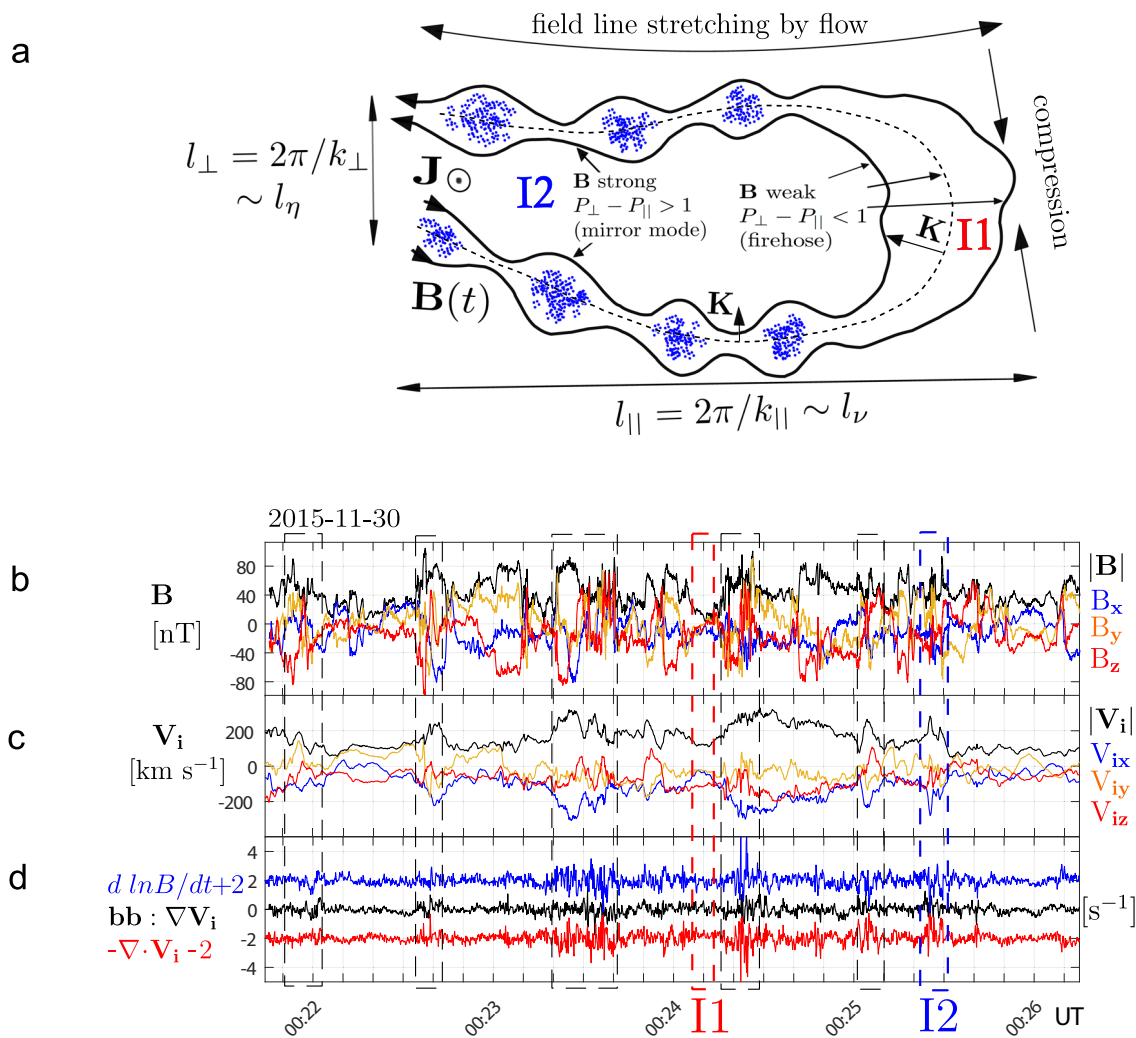

Fig. 1 | Cartoon of dynamo physics versus MMS measurements in the magnetosheath. a The expected stretched-folded magnetic field topology. Magnetic field lines (black) are advected and distorted by the plasma flow, producing stretched "arms" and strongly curved "bends." At a bend, the magnetic field **B** is weak, the curvature ($K$) is large, and the pressure anisotropy $P_\perp - P_\parallel < 0$ makes the plasma firehose unstable (example: subinterval I1, red dashed box, in the bottom panels). Along the stretched arms, the field is strong, $K$ is small, and ($P_\perp - P_\parallel > 0$) favors mirror mode instability (example: subinterval I2, blue dashed box, in the bottom panels). Blue clusters of dots indicate particles trapped in magnetic bottles associated with mirror structures. The length scale of the stretched field ($l_\parallel$) is larger than the length scale of the folded field in the perpendicular direction ($l_\perp$). These scales are comparable to viscous ($l_\nu$) or resistive scales ($l_\eta$), respectively. MMS1 measurements. Magnetic field magnitude and components (**b**) and ion velocity magnitude and components (**c**). **d** Tetrahedron terms in Eq. 1. $d \ln B/dt$ is the sum of **bb**:$\nabla$**V$_i$** and $-\nabla \cdot$**V$_i$**. The entire time interval is used for statistical analysis. The vertical red and blue dashed boxes mark subintervals I1 and I2, which are examined in detail as case studies. The vertical black dashed boxes indicate additional subintervals where enhanced MMS magnetic field and velocity fluctuations are well correlated with the calculated fluctuations of the terms in Eq. 1.

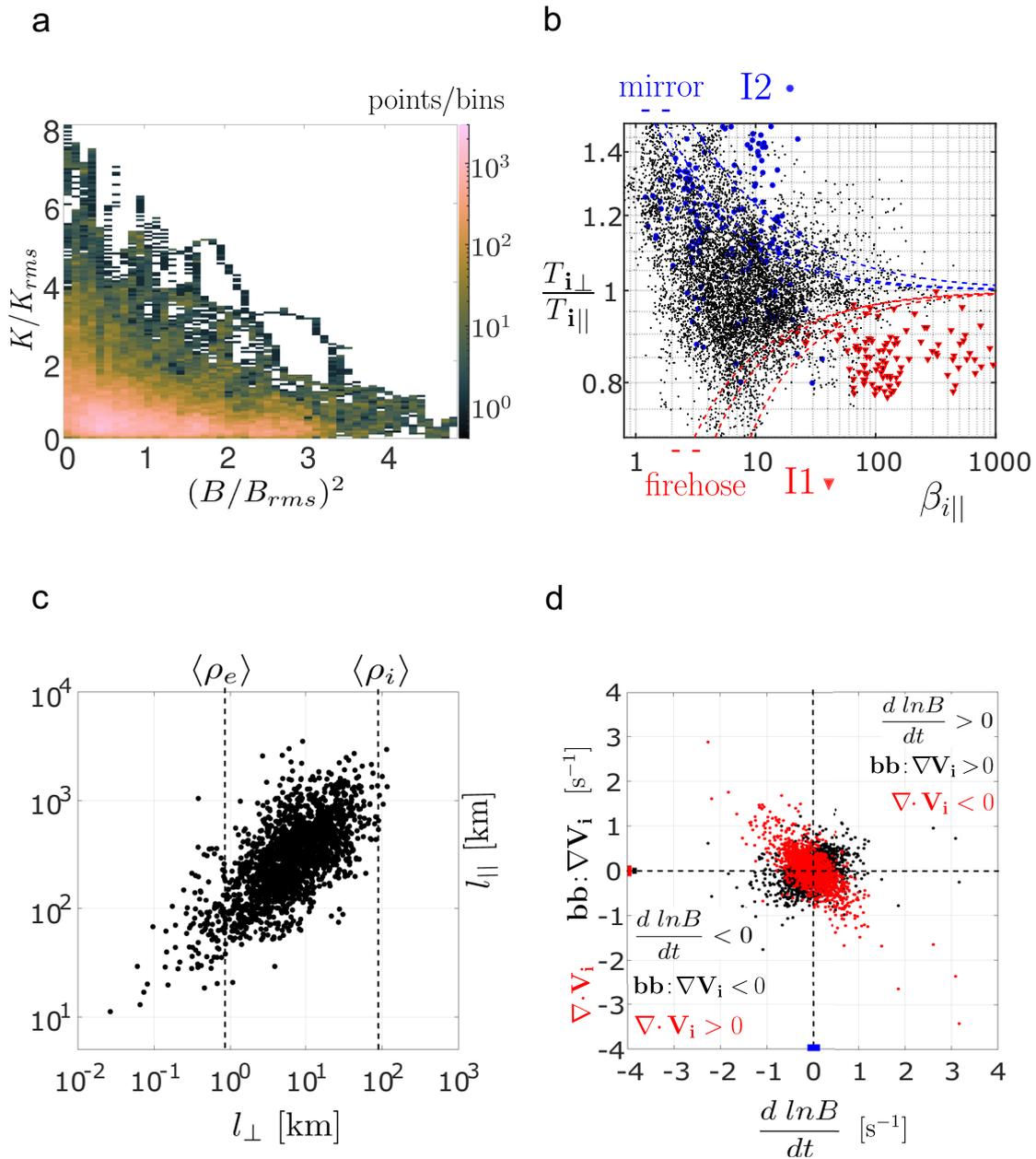

Fig. 2 | Statistical SSD signatures in MMS observations for the entire interval shown in Fig. 1 on 30 November 2015. a Anticorrelation between normalized and squared magnetic field magnitude and normalized curvature from the tetrahedron. The color code corresponds to the number of points per bin. b Parallel ion plasma beta versus the ratio of perpendicular to parallel ion temperatures for MMS1-4 spacecraft. Points corresponding to time intervals I1 and I2 are highlighted by red triangles and blue circles, respectively. The blue dashed and the red dashed curves correspond to the given instability thresholds of different growth rates[33]. c Scatterplot of perpendicular versus parallel length scales. The average ion ($\langle\rho_i\rangle$) and electron ($\langle\rho_e\rangle$) gyroradii are shown by vertical dashed lines. d The contribution of signed stretching and compression terms (rhs in Eq. 1) to the temporal changes of the magnetic field (lhs in Eq. 1, the sum of terms on the rhs).

e.g., ref. 6:

$$\frac{d\ln B}{dt} = \mathbf{bb} : \nabla \mathbf{V_i} - \nabla \cdot \mathbf{V_i} \quad (1)$$

where $d/dt \equiv \partial/\partial t + (\mathbf{V_i}\cdot\nabla)$, $B$ is the magnitude of the magnetic field, $\mathbf{V_i}$ is the ion velocity, $\mathbf{b} = \mathbf{B}/B$, $\mathbf{bb} : \nabla \mathbf{V_i} \equiv b_j b_k \nabla_j V_{i_k}$ is the double scalar product. Equation 1 describes the changes of the magnetic field through the dynamo mechanism, where the terms on the rhs are signed scalars. Only the terms on the rhs of Eq. 1 can be calculated from multi-point measurements. Direct calculation of the temporal variations on the lhs of Eq. 1 is not feasible, unless the MMS spacecraft were able to co-move with a plasma parcel in the magnetosheath. To resolve local temporal evolution, ion velocity measurements are acquired at a cadence of 150 ms, with magnetic field data resampled accordingly. The rhs quantities in Eq. 1 are evaluated at this temporal resolution, producing time series, which, combined with spacecraft separations (<20 km), correspond to sub-ion spatial scales. We consider the simplest linear problem of the kinematic dynamo when both the back reaction of the magnetic field to plasma and the magnetic diffusivity are neglected. Other processes that can influence dynamo action, such as the Hall effect[14] or cross-field particle transport[15], will be neglected as well.

Equation 1 states that any increase in the logarithm of the magnetic field magnitude, $B$, results from the stretching of **B** by field-aligned gradients of the flow velocity (**bb**:$\nabla$**V**$_i$ > 0) or compressions ($-\nabla \cdot$**V**$_i$ > 0). The reduction of the magnetic field is associated with the same terms but of the opposite signs.

The spatial gradients on the right-hand side of Eq. 1 will be estimated by using the tetrahedron high-resolution MMS data ("Multi-point estimations of gradients," in "Methods"). The distances between MMS spacecraft are less than 20 kms, corresponding to sub-ion, near electron scales in the magnetosheath. This allows in-situ observations of the turbulent SSD, as described by Eq. 1. The concept of SSDs concerns the generation or amplification of magnetic fields at scales comparable to, or smaller than, the characteristic scales of turbulent flows[16–20].

It is anticipated how the terms in Eq. 1 work. Stretchings and compressions by random velocity shears in three-dimensional plasma turbulence can twist and fold the magnetic field lines, changing their topology[17,18]. As a result, the magnetic field is weaker at the curved folds and stronger along the stretched straight field lines, leading to the anticorrelation between the magnetic field magnitude $B$ and the field line curvature $K = |(\mathbf{b}\cdot\nabla)\mathbf{b}|$[18,21]. The stretched-folded geometry of the magnetic field can also be described by characteristic wavenumbers along ($k_\parallel$) and across ($k_\perp$) **B**[18] ("Characteristic wavenumbers," in "Methods").

Although 3D turbulence can generate stretched-folded structures, the growth of B in Eq. 1 can be prevented by the conservation of the first adiabatic invariant $\mu$. The conservation of $\langle\mu\rangle$ significantly reduces the growth of the magnetic field, specifically in collisionless plasmas[22–24]. Therefore, dynamo action can effectively amplify the magnetic field when the adiabatic invariance of $\langle\mu\rangle$ is broken. Pressure-anisotropy instabilities, which can be generated by dynamo terms in Eq. 1, can break adiabatic invariance. By neglecting heat flux, compressions, and collisions, the pressure anisotropy can be calculated from[6,25,26]

$$\frac{d(P_\perp - P_\parallel)}{dt} \approx (P_\perp + 2P_\parallel)\frac{d\ln B}{dt} \quad (2)$$

where $P_\perp$ and $P_\parallel$ are pressures perpendicular and parallel to the magnetic field, respectively. According to Eqs. 1 and 2, $P_\perp > P_\parallel$ develops where $d\ln B/dt > 0$, that is, where the stretched and compressed magnetic field increases. On the other hand, $P_\perp < P_\parallel$ is associated with $d\ln B/dt < 0$, indicating a decreasing, weak magnetic field that eventually can be folded because of the small magnetic stress. In collisionless plasmas, effective collisions can appear as a result of pressure-anisotropy-driven kinetic instabilities associated with field fluctuations, which may scatter particles, breaking the $\mu$ invariance[23,27].

All the SSD-related physics introduced above is illustrated in Fig. 1a. The cartoon shows the stretched, compressed, and folded field lines with characteristic length scales of $l_\perp$ and $l_\parallel$ ("Characteristic wavenumbers," in "Methods"). The magnetic field curvature $K$ is large at curved folds where $B$ is weaker ($d\ln B/dt < 0$) and firehose instability develops. $K$ is small at stretched field lines where $B$ is strong ($d\ln B/dt > 0$), and mirror instability develops. The blue dots symbolically show the particles trapped in magnetic bottles associated with the mirror mode. The magnetic field directional changes can form current sheets associated with current density **J**, potentially reconnecting when $l_\perp$ becomes small. It is expected that in 3D turbulence, the flow-magnetic field interactions can occur in any direction[17,18], resulting in interwoven and intermittent fluctuating fields near ion-electron scales[8].

We select a time interval lasting over 4 min, during which the spacecraft samples convected plasma volumes with an along-trajectory extent of approximately 40,000 km. This distance represents a substantial fraction of the dayside magnetosheath width, assuming a convection speed of 200 kms$^{-1}$. During the interval, strong plasma turbulence was observed[28] and plasma $\beta$ occasionally and substantially exceeded solar wind values ($\beta \gg 1$), promoting pressure-anisotropy instabilities and enabling dynamo action. Field and plasma data are downloaded from the MMS archive in geocentric solar ecliptic (GSE) coordinates. Figure 1b, c shows MMS1 data obtained on 2015-11-30 between 00:21:45 and 00:26:43 UT, when the spacecraft was located in the strongly turbulent magnetosheath, downstream of a quasi-parallel shock near the magnetopause[29,30]. Figure 1d shows the dynamo terms estimated from the tetrahedral configuration. Dashed vertical boxes indicate several subintervals where the measured magnetic field and velocity fluctuations are strongly correlated with the temporal variations of terms in Eq. 1, indicating that the model adapted here is not an oversimplification. The first two terms on the right-hand side of Eq. 1 and the resulting temporal change $d\ln B/dt$ fluctuate around zero. For better visibility, two of them are shifted by values of $\pm 2$. It is noteworthy that the fluctuations within the highlighted boxes are not causally connected ("Error analysis of intermittent dynamo terms" in "Methods"). MMS observations will be used to assess whether the SSD mechanisms outlined in Fig. 1a are present within the magnetosheath. We report statistical results over the full time interval alongside detailed analyses of representative subintervals I1 and I2, marked by the vertical blue dashed boxes in Fig. 1b–d.

## Statistical results

The concept of stretched and folded magnetic fields is a cornerstone for understanding the interplay between plasma flows and magnetic topology. Figure 2a illustrates the anticorrelation between the square of the normalized magnetic field magnitude from MMS1-4 (resampled to tetrahedron barycenter) and the normalized magnetic field curvature from the tetrahedron, with root-mean-square (rms) values used for normalization. The anticorrelation, corresponding to stretched-folded field geometry, has already been identified in dynamo simulations[18], turbulence simulations[31], and magnetosheath data intervals[32]. The lack of perfect anticorrelations can be explained by the intermittency of magnetic field line structures[18] or by very different sizes of folds and MMS tetrahedron configuration. The spacecraft can also cross regions that are neither stretched nor folded.

Now we examine the abundance of pressure anisotropies during the selected time interval, which, together with the stretched-folded magnetic field topology, play a key role in enabling dynamo action. Figure 2b shows the MMS1-4 scatterplot, black points, in the $\beta_{i\parallel}$ versus $T_{i\perp}/T_{i\parallel}$ plane with overlapped instability growth rate thresholds. The anisotropic temperatures are calculated according to "Calculation of anisotropic temperatures" in "Methods." The dashed blue curves correspond to the mirror instability thresholds, and the dashed red curves correspond to the firehose instability thresholds for different growth rates[33–35]. The points in "unstable regions," that is, outside of the instability thresholds in Fig. 2b, are associated with mirror mode ($T_{i\perp}/T_{i\parallel} > 1$) or firehose instabilities ($T_{i\perp}/T_{i\parallel} < 1$), where wave modes are expected to grow, returning the plasma to "stable region" inside the thresholds. Temperature anisotropy values for the selected short intervals in Fig. 1b are highlighted by blue circles (time interval I2, mirror mode excited) and by red triangles (time interval I1, firehose instability excited). Further characterization of the magnetic field topology, based on its characteristic scales, enables comparative analysis across parametrized dynamo models. Figure 2c shows the distribution of magnetic length scales $l_\perp$ and $l_\parallel$ estimated from the characteristic wavenumbers ("Characteristic wavenumbers," in "Methods," Eqs. 3 and 4). The average ±standard deviation of the ion and electron gyroradii are 95 ± 84 km and 0.9 ± 0.63 km, respectively. The $l_\perp$ values are mainly located over the sub-ion scales. Nevertheless, only two events with reconnection signatures and short $l_\perp$ (thin current sheets) have been reported for the entire time interval[29,30]. From Fig. 2c, we deduce $l_\parallel/l_\perp = 40 \pm 20$. Given that $l_\perp \sim l_\eta \sim Pm^{-1/2}l_\nu \sim Pm^{-1/2}l_\parallel$[6], it follows that $Pm \in (400, 3600)$. Here, $Pm$ denotes the Prandtl number, a key parameter in dynamo research. A large magnetic Prandtl number ($Pm \gg 1$) in collisionless turbulence suggests that the plasma behaves similarly to a "stretch-and-fold" dynamo—a regime where magnetic field amplification is dominated by viscous-scale velocity fluctuations, even in the absence of frequent particle collisions. As a consequence of the stretching and folding process, the resulting magnetic field topology can evolve into current sheets[36].

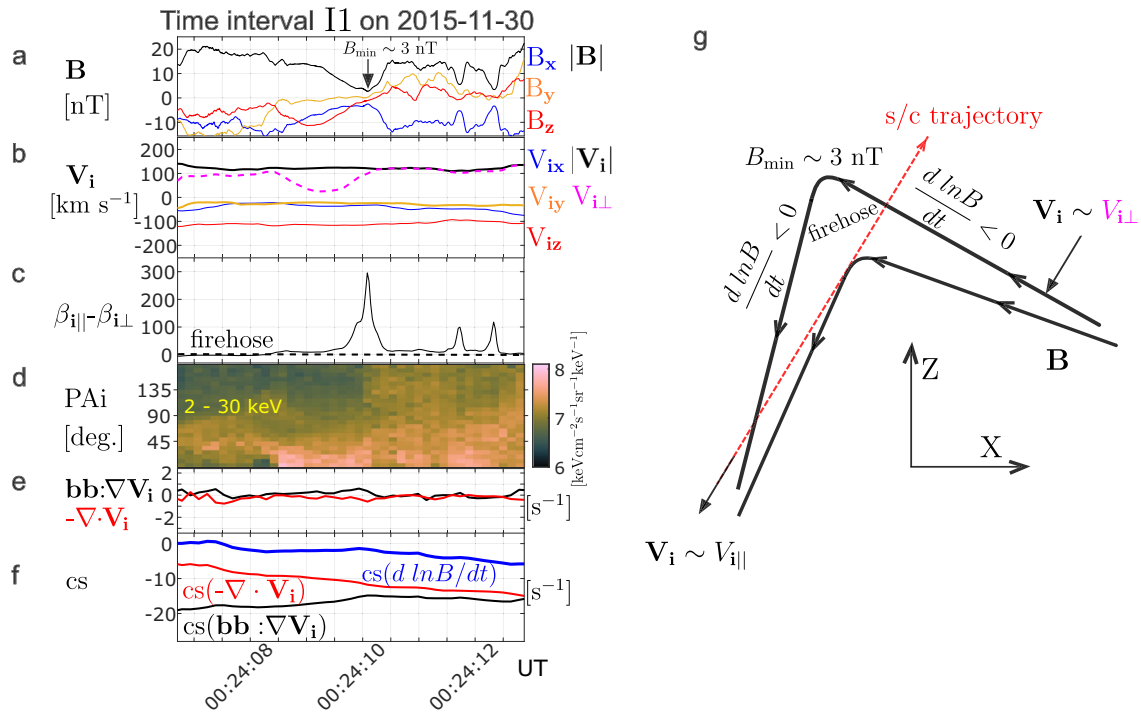

**Fig. 3 | Time interval I1: the folded magnetic field. a–d** MMS1 measurements. **a** Magnetic field magnitude and components. **b** Ion velocity magnitude, components, and the component perpendicular to **B**. **c** Firehose instability criterion: $\beta_\| - \beta_\perp > 2$. **d** Pitch angle distribution for ions. The color code is for the energy flux of particles. **e, f** Tetrahedron measurements. **e** Stretching and compressional terms from Eq. 1. **f** Cumulative sums (cs). **g** Cartoon illustrating the crossing of a magnetic fold. The figure shows how the measured field, plasma parameters, particle data, and dynamo terms (**a–f**) vary along the spacecraft trajectory (red dashed arrow in **g**). In the **X–Z** plane, the $B_z$ component changes sign, while $d \ln B/dt$ decreases on both wings toward $B_{min} \sim 3$ nT, where the field is weak and susceptible to folding. Additional signatures of the folded field include a negative $B_x$ and the occurrence of the firehose instability. Although $V_i$ is unchanged, $V_{i\perp}$ differs across the two wings (**b, g**), indicating that spatial gradients in $V_{i\perp}$ support further folding of the structure.

Figure 2d displays the relationship between different terms from Eq. 1. The scatterplot between $d \ln B/dt$ and $\mathbf{bb}:\nabla \mathbf{V_i}$ is shown with black points, while the scatterplot between $d \ln B/dt$ and $\nabla \cdot \mathbf{V_i}$ is shown with red points, in the same figure. We remind that $d \ln B/dt$ cannot be directly calculated, but is entirely determined by the quantities on the rhs of Eq. 1. The same Y axis is used for the stretching and compression terms. It is indicated in the right half of Fig. 2d that a positive stretching term (reduction of the volume with field lines) and a negative compression term (converging flows) lead to the growth of the magnetic field. At the same time, as indicated in the left half of the figure, the decrease of the magnetic field is associated with a negative stretching term (dilation of the volume with field lines) and a positive compression term (diverging flows). Notice that a positive value of $-\nabla \cdot \mathbf{V_i}$ is associated with an increasing magnetic field. The average errors of different quantities ("Error analysis of intermittent dynamo terms," in "Methods") in Fig. 2d are indicated near the zero abscissa and ordinate values, respectively. The horizontal blue line is the average error for $d \ln B/dt$ (±0.1 s$^{-1}$), while the vertical black and red lines represent the average errors for $\mathbf{bb}:\nabla \mathbf{V_i}$ (±0.05 s$^{-1}$) and $\nabla \cdot \mathbf{V_i}$ (±0.09 s$^{-1}$), respectively.

These statistical results indicate that the expected signatures of turbulent dynamo activity—namely, magnetic topology changes and pressure-anisotropy-driven instabilities—are abundantly present in the considered part of the magnetosheath. Moreover, the stretching and compressive terms (rhs of Eq. 1) necessarily drive temporal variations in the magnetic field (lhs of Eq. 1).

## Case studies (intervals I1 and I2)

To gain deeper insight into the SSD dynamics described above and depicted in Fig. 1a, we examine two representative time intervals, I1 and I2.

First, we present a case of a folded magnetic field exhibiting the expected signatures of dynamo action along the spacecraft trajectory. These signatures include a weakened magnetic field at a magnetic fold, associated with firehose instability, and a gradient in the velocity component perpendicular to the magnetic field across the folded topology. The latter acts to drive oppositely directed field lines closer together, facilitating the formation of a thinner current sheet.

Figure 3a–d shows MMS1 data, and Fig. 3e, f shows the terms in Eq. 1 and their cumulative sums (cs) during interval I1. The cs($d \ln B/dt$) curve is shifted for better visibility. A cs value for a given variable at any time is the sum of all previous values of the considered variable. The cumulative aggregation of decisively positive, negative, or zero values over a time interval leads to increasing, decreasing, or unchanging cs curves, respectively. While the variables in Fig. 3e intermittently fluctuate, their cs' show more clearly when these terms are changing or are relatively constant in time. As it is explained in "Error analysis of intermittent dynamo terms" (in "Methods"), the increasing or decreasing parts of cs' can be treated independently. The errors of changing cs' are about 10%. Now we compare the data with the cartoon in Fig. 3g showing the spacecraft crossing of a folded magnetic field. Figure 3a shows that $B$ is decreasing from the beginning of the time interval, reaching minimum values near 00:24:10 UT, then fluctuating

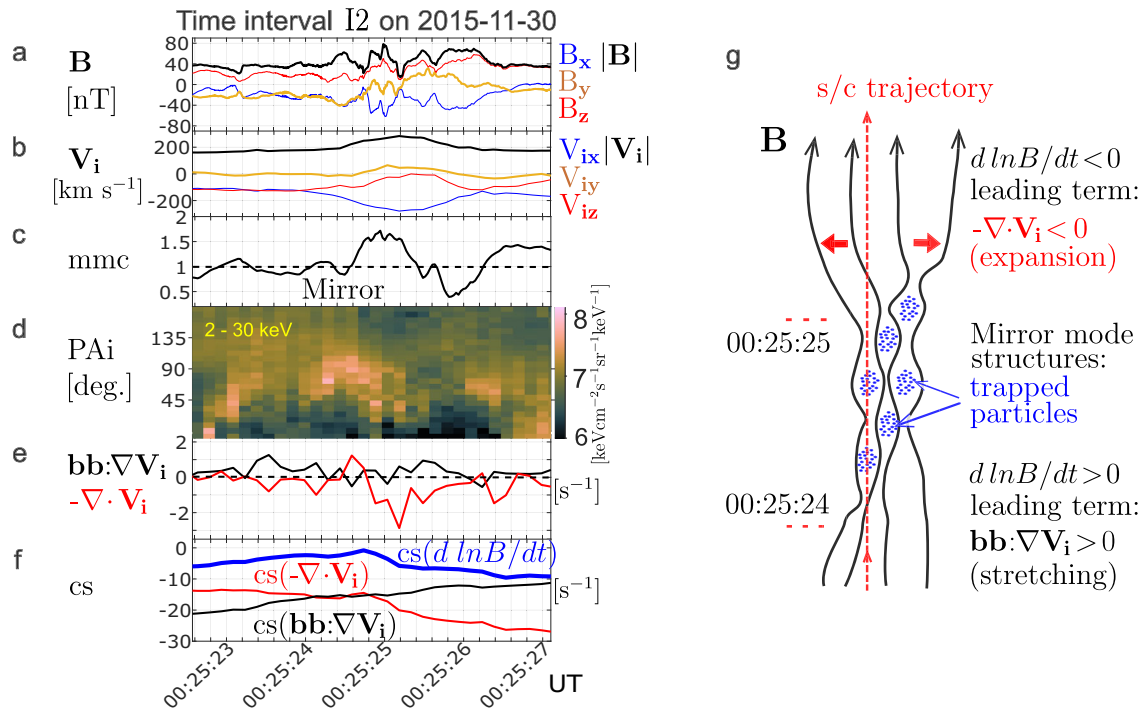

**Fig. 4 | Time interval I2: mirror mode instability and the stretched-expanded magnetic field. a**–**d** MMS1 measurements. **a** Magnetic field magnitude and components. **b** Ion velocity magnitude and components. **c** Mirror mode instability criterion: mmc > 1. **d** Pitch angle distribution for ions. The color code is for the energy flux of particles. **e**, **f** Tetrahedron measurements. **e** Stretching and compressional terms from Eq. 1. **f** Cumulative sums (cs). **g** Cartoon illustrating the crossing of an increasing magnetic field associated with mirror instability and a decreasing magnetic field through expansion. The figure shows how the measured parameters in (**a**–**f**) vary along the spacecraft trajectory (red dashed arrow in **g**). The dashed red horizontal lines indicate approximate timestamps depicted in the cartoon.

and finally increasing at the end of the time interval. It can be interpreted as a decrease of B towards its minimum from both sides. At the same time, cs($d \ln B/dt$) decreases or stagnates. During the time interval I1 cs($d \ln B/dt$) decreases by $-6 \pm 0.7\,s^{-1}$ ("Error analysis of intermittent dynamo terms" in "Methods"). $B_y$ and $B_z$ are first negative, then positive; $B_x$ is always negative. The magnetic field topology is also shown in the X–Z plane in the cartoon (in the X–Y plane it is similar). At the fold $|B_x| \sim 3$ [nT] $\sim B_{min}$. At the same time, firehose instability occurs over the fold. Figure 3c demonstrates the fulfillment of the firehose instability criterion, $\beta_{i\|} - \beta_{i\perp} > 2^{37}$, which is associated with an enhanced parallel flux of ions (observed at small PAi angles in Fig. 3d) and temperature anisotropy, $T_{i\perp}/T_{i\|} < 1$, within a high $\beta_{i\|}$ environment (Fig. 2b). The occurrence of firehose instability, the decreasing $d \ln B/dt$ and the negative and dominating Bx in between oppositely oriented By and Bz components are strongly indicating that the spacecraft are crossing a folded magnetic topology as it is depicted in the cartoon. Moreover, $V_i$ shows no changes; however, due to the directional change of **B**, the velocity component perpendicular to local **B**, $V_{i\perp}$ is significantly different over the two wings of the magnetic structure (Fig. 3b, g). When the magnetic field becomes locally weaker due to the combined action of stretching and compressive terms, the magnetic field can be folded by the spatial gradients of the perpendicular velocity components $V_{i\perp}$.

We now present a case in which dynamo action strengthens the magnetic field along the spacecraft trajectory, coinciding with the occurrence of mirror mode instability. In this case, plasma volumes are dominated by stretching as B increases. Figure 4a–d shows MMS1 data, and Fig. 4e, f shows the terms in Eq. 1 and their cs' during interval I2. Around 00:25:25 UT, the magnetic field exhibits strong and rapid fluctuations and an increase in magnitude from 40 to 80 [nT]. The increasing and fluctuating magnetic field (**B**) is accompanied by directional and magnitude variations of the ion velocity ($V_i$). At the same time, the mirror mode instability criterion,

mmc = $T_{i\perp}/T_{i\|} - 1/\beta_{i\perp} > 1^{38}$, is satisfied. This instability is linked to an enhanced perpendicular flux of ions (observed at approximately 90° pitch angles in Fig. 4d) and temperature anisotropy ($T_{i\perp}/T_{i\|} > 1$) within a high $\beta_{i\|}$ environment (Fig. 2b). The stretching term cs(**bb**:$\nabla$**V**$_i$) (black curve) increases during the whole time interval B. cs($d \ln B/dt$) increases by $6 \pm 0.5\,s^{-1}$, approximately until 00:25:25 UT. Afterward, cs($-\nabla \cdot$**V**$_i$) starts decreasing faster, with some varying rate, than the cumulative stretching term increases. As a consequence, cs($d \ln B/dt$) decreases by $-9 \pm 0.9$ [1/s]. Figure 4g summarizes what is seen along the spacecraft (s/c) trajectory. First, the magnetic field increases ($d \ln B/dt > 0$) due to the strong positive stretching term in Eq. 1. As a consequence, the mirror mode structures trap or scatter particles, breaking the adiabatic invariance. Afterward, the leading term in Eq. 1 is ($-\nabla \cdot$**V**$_i$) < 0, indicating that an expansion results in $d \ln B/dt < 0$.

The case studies, reinforcing the statistical findings, provide compelling evidence for the existence of sub-ion-scale turbulent dynamo mechanisms in the collisionless magnetosheath.

## Discussion

The experimental results show that by combining single-point and multi-point measurements of plasma and magnetic fields, it is possible to identify local turbulent dynamo activity—responsible for changes in magnetic field strength—and associated pressure-anisotropy instabilities within the terrestrial magnetosheath. It is anticipated that SSD observations in the magnetosheath can be compared to different dynamo simulation regimes on the basis of dimensionless quantities, such as the Prandtl number $Pm$. In natural systems, $Pm$ can reach very large values (e.g., $\sim 10^{25}$ in hot inter-galaxy-cluster medium), very low values (e.g., $\sim 10^{-5}$ in the Earth's core), or values near 1 (e.g., near surface turbulent solar dynamo). The accessible range of $Pm$ in numerical simulations or laboratory dynamo experiments is $0.1 < Pm < 100^{27}$. Previous studies, based on the estimation of energy exchanges in numerical simulations and observations, found that $Pm$ in the magnetosheath is approximately 1[39]. However, properties of SSD driving turbulence can strongly depend on position[40–43] and physical parameters such as compressibility[44,45], plasma $\beta$, and instability thresholds[46] or upstream solar wind structures and their interaction with the bow shock[47]. It is plausible that effective viscosity (e.g., due to wave-particle interactions) and magnetic diffusivity (e.g., through current sheet formation) vary across space and time, which could lead to broad fluctuations in $Pm$. The selected interval in this paper lies near the magnetopause within a quasi-parallel, highly turbulent magnetosheath, under stable upstream conditions. For this time interval, we found that $Pm$ is two to three orders of magnitude larger than found previously[39], indicating that the magnetosheath can be a real test-bed for a broad range of turbulent dynamo theories and simulations. For $Pm > 1$, the characteristic field reversals can form potentially reconnecting current sheets, typically near the effective resistive scale[36]. Thin current sheets are ubiquitous in magnetosheath turbulence; however, not all of them undergo reconnection.[48]. Specifically, during the interval shown in Fig. 1, a number of current sheets were observed[28], of which only two exhibited clear signatures of magnetic reconnection[29,30]. While 3D turbulence in the magnetosheath can stretch and fold magnetic field lines into thin, reconnecting current sheets, tracing the temporal evolution of this process—particularly the transition from dynamo action to reconnection—remains challenging. This difficulty arises because the MMS spacecraft cannot co-move with the plasma parcel undergoing the temporal transformation. Consequently, our analysis focuses on the instantaneous signatures of dynamo terms alone, rather than their temporal evolution. Hybrid-kinetic simulations of current sheet formation, however, demonstrate that a thinning current sheet is associated with the onset of mirror instability at sufficiently high plasma $\beta$[49]. The simulation results are interpreted in relation to the stretched and folded magnetic field structures, as well as dynamo action occurring in the presence of temperature anisotropy.

In a broader context, high-resolution MMS measurements provide substantial evidence that small-scale magnetic reconnection and turbulence represent distinct facets of multi-scale energy transfer and conversion in collisionless plasmas[50]. Although this paper focuses on identifying SSD mechanisms, it is important to recognize that other energy conversion pathways exist at sub-ion scales in 3D collisionless plasma turbulence. In particular, turbulent interactions can facilitate energy transfer between bulk flow and thermal energy via pressure–velocity strain coupling, as well as through field-particle interactions that contribute to electromagnetic energy changes and particle acceleration[31,51]. Focusing on the SSD problem thus inherently advances our understanding of collisionless turbulence at scales approaching the kinetic dissipation range—a regime long regarded as critically important for unraveling the fundamental processes that govern energy transfer in plasmas. SSD action requires strong velocity gradients aligned with the magnetic field, as well as compressive structures. Because of this, it is reasonable to assume that specific plasma conditions and geometric configurations play a role in regulating energy conversions[52]. These factors may influence whether the free energy in the velocity gradient field is primarily driving SSD or is converted into heat through pressure-strain interactions or is channeled into electromagnetic energy variations and particle acceleration via field-particle interactions. The extent to which these pathways compete or coexist in turbulent plasma remains an open question. To disentangle the multi-scale energy conversion processes—spanning dynamo action across varying magnetic Prandtl numbers and mechanisms such as pressure-strain and field-particle interactions—a comprehensive statistical framework is required, particularly under changing solar wind conditions. Such analysis is key to identifying which classes of space and astrophysical dynamos may be accessible in this natural laboratory. At the same time, a wider statistical analysis would offer a pathway to assess the role of turbulent dynamo processes in the broader context of collisionless turbulence, particularly at scales approaching kinetic dissipation. Understanding these dynamics remains a central challenge in plasma physics.

## Methods

### Multi-point estimation of gradients

The MMS spacecraft fly in a tetrahedron configuration (a pyramid-like formation), with each spacecraft having the same instrumentation. For the estimation of gradients, the method of linear interpolation combined with Ampère's and Gauss's integral theorems is used. In this approach, the position vectors of the spacecraft (the vertices of the tetrahedron) are used to create reciprocal vectors, which are proportional to the area of the face of the tetrahedron opposite to a vertex and inversely proportional to the volume of the tetrahedron. Then the gradients, divergences, and curls of scalar and vector fields can be calculated[53].

### Characteristic wavenumbers

The stretched-folded geometry of the magnetic field can be described by characteristic wavenumbers along ($k_\parallel$) and across ($k_\perp$) $\mathbf{B}$[18]

$$k_\parallel = \left( \frac{\langle |\mathbf{B} \cdot \nabla \mathbf{B}|^2 \rangle}{\langle B^4 \rangle} \right)^{1/2} \quad (3)$$

$$k_\perp = \left( \frac{\langle |\mathbf{B} \times \mathbf{J}|^2 \rangle}{\langle B^4 \rangle} \right)^{1/2} \quad (4)$$

where $\mathbf{J} = \mu_0^{-1}(\nabla \times \mathbf{B})$, $\mu_0$ is the permeability of free space. We note that instead of the averaged values of variables ($\langle variable \rangle$), the local quantities are calculated in each time. The characteristic length scales can be calculated through $l_\parallel = 2\pi k_\parallel^{-1}$ and $l_\perp = 2\pi k_\perp^{-1}$.

### Calculation of anisotropic temperatures

The parallel and perpendicular to the magnetic field ion temperatures are obtained by rotating the temperature tensor into field-aligned coordinates, where the background magnetic field was calculated locally at the sampling rate of the particle measurement. The perpendicular component is calculated by taking the average of the two components perpendicular to the magnetic field plane.

### Error analysis of intermittent dynamo terms

At any given time, the local errors of the terms in Eq. 1, which include spatial gradients, are estimated from tetrahedron measurements using the Monte Carlo method described in ref. 54. The components of the magnetic field and velocity vectors are perturbed with random fluctuations having zero means and standard deviations as the examined physical quantities. This procedure is repeated, generating one hundred perturbed realizations of the time series. The local means and standard deviations of the terms in Eq. 1 are calculated from the perturbed ensemble of time series. The averaged standard deviations for the whole time interval are shown in Fig. 2d near the zero abscissa and ordinate values, respectively. However, there are reasons why the error bars are not shown in Fig. 1d. First of all, in turbulent plasmas, energy conversions are intermittent and concentrated at or near coherent structures, such as current sheets, vortices[51] or their groups[55]. The intermittently enhanced or diminished fluctuations are clearly seen in Fig. 1d. Moreover, the typical correlation lengths of magnetosheath fluctuations are short[40]. As a result, the energy conversions at different locations, or within the highlighted boxes in Fig. 1b–d, are not causally connected. Consequently, the errors in the cumulative sums (cs) do not increase over the entire time interval shown in Fig. 1. Instead, they can be calculated for finite durations (or finite plasma volumes) when enhanced energy exchanges occur. It is worth noting that the cumulative errors within the selected time intervals are comparable to the thickness of the cs lines in Figs. 3f and 4f. Therefore, error bars are not displayed.

## Data availability

MMS data are publicly available at https://lasp.colorado.edu/mms/sdc/public/about/browse-wrapper/, where in the folders mms1-4, the magnetic field data are under /fgm/brst/l2/2015/11/30/, and the ion moments data, including velocity, pressure, and temperature, are under fpi/brst/l2/dis-moms/2015/11/30/, and pitch angle data are under fpi/brst/l2/dis-dist/2015/11/30/. Source data are provided with this paper.

## Code availability

The estimation of spatial gradients from multi-spacecraft tetrahedron measurements is described in ref. 53 implemented in the software package irfu-matlab, https://github.com/irfu/irfu-matlab. Irfu MATLAB can also be downloaded from Zenodo[56]. The following functions were applied: c_4_grad.m for calculating divergences, gradients, and curls; mms.rotate_tensor.m for calculating parallel and perpendicular components of the temperature tensor; irf_resamp.m for resampling a quantity onto a timeline of another quantity. The scripts for reproducing the figures are uploaded to Zenodo https://doi.org/10.5281/zenodo.17780770[57].

## Acknowledgements
The authors thank the MMS team and the MMS Science Data Center for providing high-quality data for this study. This research was funded in part by the Austrian Science Fund (FWF) under 10.55776/PAT9232923. For the purpose of Open-Access, the author has applied a CC BY public copyright licence to any Author Accepted Manuscript version arising from this submission; Z.V. and P.A.B. were also funded by the Austrian Science Fund (FWF) P 37265-N; Y.N. was funded by the German Science Foundation under project number 535057280; E.Y. was funded by the Swedish National Space Agency (SNSA) under grant 2020-00192. The work of C.S.W. was funded by the Austrian Science Fund (FWF) 10.55776/P35954. L.S.-V. was supported by the Swedish Research Council (VR) Research Grant N. 2022-03352, by the project 2022KL38BK. L.S.-V. also acknowledges the projects PRIN/PNRR-H53D23011020001 and PRIN-2022KL38BK, supported by the Italian Ministry of University and Research.

## Author contributions
Z.V. initiated this study, did the analysis, and wrote the paper. O.W.R., Y.N., and P.A.B. gave the initial idea of the LSD in the turbulent solar wind; E.Y. contributed to data preparation and analysis, writing and editing the manuscript. R.N., A.S., D.S., M.W., C.L.S.W., and A.V. contributed to the analysis methods and helped to edit the paper. L.S.V. contributed to the idea of SSD in a turbulent plasma environment. Á.K. contributed conceptual guidance on the turbulent magnetosheath that strengthened the framing of the study.

## Competing interests
The authors declare no competing interests.

## Consent to publish
All authors give permission to publish the work.

## Additional information
**Supplementary information** The online version contains supplementary material available at https://doi.org/10.1038/s41467-026-69469-y.

**Correspondence** and requests for materials should be addressed to Zoltán. Vörös.

**Peer review information** *Nature Communications* thanks the anonymous reviewers for their contribution to the peer review of this work. A peer review file is available.

**Reprints and permissions information** is available at http://www.nature.com/reprints

**Publisher's note** Springer Nature remains neutral with regard to jurisdictional claims in published maps and institutional affiliations.